\documentclass[twocolumn,showpacs,preprintnumbers,amsmath,amssymb,eqsecnum]{revtex4}
\usepackage{graphicx}
\begin{document}
\preprint{RCA-02-01}
\title{Stable accelerating universe with no hair}

\author{Pedro F. Gonz\'{a}lez-D\'{\i}az}
\affiliation{Centro de F\'{\i}sica ``Miguel Catal\'{a}n'', Instituto de
Matem\'{a}ticas y F\'{\i}sica Fundamental,\\ Consejo Superior de
Investigaciones Cient\'{\i}ficas, Serrano 121, 28006 Madrid (SPAIN)}
\date{\today}

\begin{abstract}
After reviewing the main characteristics of the spacetime of
accelerating universes driven by a quintessence scalar field with
constant equation of state $\omega$, we investigate in this paper
the classical stability of such spaces to cosmological
perturbations, particularizing in the case of a closed geometry
and equation of state $\omega=-2/3$. We conclude that this space
is classically stable and conjecture that accelerating universes
driven by quintessential fields have "no-hair".
\end{abstract}


\maketitle

\section{Introduction}

Knowing the fate of the universe has always possibly been the main
goal of cosmology. The speculation on what the future of the
universe might be has now become even more complicated. Following
recent observations in supernovas Ia [1] a real revolution has
stirred cosmology, leading to a plethora of new views, concepts
and finally to an emerging entirely new cosmological scenario
which some like to call the {\it New standard Cosmology} [2].
Critically pivotal of such a scenario is the already
overwhelmingly accepted view that the expansion of the present
universe is accelerating [3]. Whether this acceleration would
continue forever or it will stop to recover decelerating expansion
is still not known. The key ingredient that provides the repulsive
force required by an accelerating universe is cosmic dark energy
which, even though is not directly detectable, should make nearly
the seventy percent of the total energy content in the universe.
So far, two main candidates have been considered for dark energy:
a positive cosmological constant [4] and the so-called
quintessence field, a slowly-varying scalar field with negative
pressure which only recently (in cosmic time) has started to
dominate over all forms of matter [5,6]. Several compelling
arguments have been advanced [5-7] in favor of cosmic quintessence
with respect to the cosmological constant.

It has recently been argued however [8-10] that if accelerating
expansion continues eternally, then a future event horizon will
inexorably form and this would represent the demise for any
mathematically consistent formulation of quantum gravity and
string theory, both in the case that the universe be now dominated
by a positive cosmological constant or by a cosmic quintessential
field with constant [5] or time-dependent [6] equation of state.
This possibly is one of the greatest challenges ever posed to
theoretical physics. The formation of a future event horizon,
which would make it impossible to construct any consistent
S-matrix for string theory, is an unavoidable consequence from the
causal future development of any initial surface in the
accelerating universe if quantum coherence is to be preserved.
Nevertheless, there exist solutions to the static Einstein
equations for an eternally accelerating universe endowed with a
quintessence scalar field with constant equation of state which do
not show any event horizon [11]. This is made possible because
such solutions possess a Kleinian signature that can lead to
allowance of world lines traveling backward in time. Relative to
the case of a positive cosmological constant, this may become one
of the strongest arguments in favor of quintessence.

Along the evolution of the universe from its quantum era, tracking
models can actually be represented [12] as a succession of
$\omega=p/\rho$=Const. domains, ranging from $\omega=+1$ to
$\omega=-1$, separated by abrupt jumps and followed by the present
accelerating expansion with $-1/3>\omega>-1$, which evolves from
an attractor solution [6,12,13] and may or may not be
characterized by a constant $\omega$. However, it seems to be a
good enough approximation to also represent the present period of
dark-energy dominated evolution by a quintessence model with
$\omega$=Const. Therefore, investigating the spacetime of
eternally accelerating universes whose expansion is driven by a
quintessence scalar field with constant equation of state and does
not show future event horizon appears to be of interest. This will
be done in the present paper, specializing in the case
$\omega=-2/3$. We shall also deal with the issue of the stability
of this eternally accelerating universe by considering the
cosmological Lifshitz-Khalatnikov perturbations [14] on it. We
check that the considered spacetime is stable and advance the
conjecture that an eternally accelerating closed universe has
no-hair.

The paper can be outlined as follows. In Sec. II we review the
solutions to the Einstein equations corresponding to a
quintessence scalar field with constant $\omega$, minimally
coupled to Hilbert-Einstein gravity, both in the static and
cosmological cases, for the whole range of state equations that
covers all possible universes with accelerating expansion. We
particularize in the case $\omega=-2/3$ whose FRW metric is
considered in some detail for the different geometries of the
universe. The Lifshitz-Khalatnikov cosmic perturbations of this
spacetime are studied in detail in Sec. III and it is checked that
the spacetime is stable to them. We also advance the conjecture
that an eternally accelerating closed universe has no-hair,
discussing it by comparing with the purely de Sitter space.
Finally we conclude and add some further remarks in Sec. IV.

\section{The spacetime of an accelerating universe}

The spacetime of the accelerating Friedmann-Robertson-Walker (FRW)
universe endowed with a quintessence scalar field with constant
state equation parameter $-1/3>\omega >-1$ corresponds to
maximally symmetric spaces with negative spacetime curvature which
are solutions of the Einstein equations. Since the conservation
law for a generic quintessence scalar field should be taken to be
$\alpha\rho=a(t)^{-3(1+\omega)}$ (where $\alpha$ is an arbitrary
integration constant, $\rho$ is the energy density of the
quintessence field and $a(t)$ is the time-dependent scale factor),
the Friedmann equations for the quintessential accelerating
universe are [15]
\begin{equation}
a^{2+3\omega}\ddot{a}+\frac{4\pi G(1+3\omega)}{3\alpha}=0
\end{equation}
\begin{equation}
\frac{\dot{a}^2}{a^2}\equiv H^2= \frac{8\pi
G}{3\alpha}a^{-3(1+\omega)}\pm \frac{1}{R^2 a^2} ,
\end{equation}
in which $H$ is the Hubble constant, $R^{-2}$ is the spatial
curvature constant, and the overhead dot denotes differentiation
with respect to time $t$. Note that if we set $\omega=-1$ the
first term of the right-hand-side in Eq. (2.2) becomes a constant,
and we obtain then a solution for the scale factor that describes
just de Sitter space. Here we shall restrict ourselves to solve
Eqs. (2.1) and (2.2) in two particular interesting cases. When we
set $\omega=-1/3$, i.e. at the onset of the accelerating regime,
we obtain the solution [16]
\begin{equation}
a(t)=\sqrt{\frac{8\pi G}{3\alpha}-\frac{1}{R^2}}t+K_0 ,
\end{equation}
where $K_0$ is an integration constant. Whereas in the spatially
closed case with $R^{-2}=8\pi G/(3\alpha)$ the scale factor (2.3)
reduces to a simple constant that describes an Einstein static
universe, in the spatially flat and open cases, or when
$R^{-2}<8\pi G/(3\alpha)$ for closed geometry, the universe will
expand in just the uniform way.

For a constant equation of state with $\omega=-2/3$, i.e. at the
typical most interesting situation in which the universe expands
in an accelerating fashion quite adjustable to what has been
observed in recent supernova experiments [1,3], the scale factor
solution to the Friedmann equations (2.1) and (2.2) can generally
be written
\begin{equation}
a(t)=\frac{2\pi Gt^2}{3\alpha} +Kt
+\frac{3\alpha\left(K^2+R^{-2}\right)}{8\pi G} .
\end{equation}
If, without any loss of generality, we set the integration
constant $K=0$ (in what follows it will be seen that when we
re-express the scale factor in terms of the compactified time
$\eta$, the resulting solution does not explicitly depend on $K$),
then this solution described the FRW spacetime of half a
Lorentzian wormhole [17], from a throat at $t=0$, where $a=a_0=
3\alpha/(8\pi GR^2)$, to the asymptotic region at $t=+\infty$. We
furthermore notice that if for $K=0$ we allowed time $t$ to take
also on negative values from $t=0$ down to $t=-\infty$, then a
complete wormhole would be obtained. Let us next assume for a
moment that the two asymptotic regions at $t=\pm\infty$ are
identified to each other so that [18] any world lines for test
particles or light signals approaching the infinity surface at
$t=+\infty$ from $t=0$ would find themselves emerging from the
infinity surface at $t=-\infty$, toward $t=0$, and any of such
lines approaching the infinity surface at $t=-\infty$ also from
$t=0$ would emerge from the infinity surface at $t=+\infty$, again
toward $t=0$. None of these world lines had then reached any of
the infinities. Thus, if the universe can be described as a
compactified complete wormhole the way we have just described,
then light signals traveling backward in time existed and could
connect the whole spacetime in such a way that no event horizon
would form up in the future [11]. This is by no way implying the
existence in the accelerating universe of any nonchronal regions
containing closed timelike curves because for these curves to
occur in our spacetime it would be necessary that the two
asymptotic regions at $t=\pm\infty$ were also set into motion
relative to one another [18], which is not possible for the
universe.

In what follows, we shall interpret the throat of the wormhole at
$t=0$ as the latest hypersurface of the universe immediately
before the onset of the accelerating regime. This will make the
value of the constant $3\alpha/8\pi G$ very large. The three
possible geometries associated with the FRW metric that correspond
to solution (2.4) with $K\neq 0$ lead to the following expressions
for the scale factor in terms of the conformal time $\eta=\int
dt/a(t)$.

\noindent (i) Spatially closed $R^{-2}>0$
\begin{equation}
a(\eta)=\frac{3\alpha}{8\pi GR^2 \cos^2\left(\eta/(2R)\right)} ,
\end{equation}
with
\begin{equation}
\eta=2R\arctan\left[R\left(K+\frac{4\pi
Gt}{3\alpha}\right)\right],
\end{equation}

\noindent (ii) Spatially flat $R^{-2}=0$
\begin{equation}
a(\eta)=\frac{3\alpha}{2\pi G\eta^2},
\end{equation}
with
\begin{equation}
\eta=-\frac{2}{K+\frac{4\pi Gt}{3\alpha}} ,
\end{equation}

\noindent (iii) Spatially open $R^{-2}<0$
\begin{equation}
a(\eta)=-\frac{3\alpha}{8\pi G|R|^2
\cosh^2\left(\eta/(2|R|)\right)} ,
\end{equation}
with
\begin{equation}
\eta=-2|R|{\rm arctanh}\left[|R|\left(K+\frac{4\pi
Gt}{3\alpha}\right)\right].
\end{equation}
Note that in all three cases the scale factor does not explicitly
depend on the integration constant $K$.

Although particular values of the parameter $\omega$ such as
$\omega=-1/3$ and $\omega=-2/3$ would strictly make the usual
sense only for homogeneous and isotropic FRW spacetimes, if we
keep an equation of state $p=\omega\rho$ also in the case of
spacetimes with static, spherically symmetric coordinates, one can
also obtain the static metrics which correspond to the above
particular cases for a given, fixed relation between the energy
density and the metric components, much in the same way as the
static metric for de Sitter space can be derived from the Einstein
equations for static, spherically symmetric coordinates and an
equation of state $p=\omega\rho$, whenever we set $\omega=-1$.
That static de Sitter metric can directly be related with a
corresponding FRW metric for de Sitter space obtained from Eqs.
(2.1) and (2.2) also for $\omega=-1$ by means of embeddings in a
common five-dimensional hyperboloid (see Refs. [20] and [21]). We
next consider the generic metric that describes the spacetime in
static, spherically symmetric coordinates $t, r, \theta, \phi$ for
an equation of state $p=\omega\rho$ and positive spatial
curvature, corresponding to an {\it ansatz}
\[ds^2=-A(r)dt^2+B(r)dr^2+r^2 d\Omega_2^2 ,\]
where $d\Omega_2^2$ is the metric on the unit two-sphere, and
$A(r)$ and $B(r)$ are metrical coefficients depending only on the
radial coordinate $r$. Following Ref. [11], we take as the
components of the energy-momentum tensor
$T_r^r=T_{\theta}^{\theta}=T_{\phi}^{\phi}=p$ and $T_0^0=-\rho$,
that is $T_r^r=T_{\theta}^{\theta}=T_{\phi}^{\phi}=-\omega T_0^0$,
and hence we have the Einstein equations
\begin{eqnarray}
&&8\pi
G\omega\rho=\frac{1}{B}\left(\frac{1}{r^2}+\frac{A'}{rA}\right)
-\frac{1}{r^2}\nonumber\\ &&=\frac{1}{4B}\left(\frac{2A''}{A}-
\frac{(A')^2}{A^2}-\frac{2B'}{rB}-\frac{A' B'}{AB}
+\frac{2A'}{rA}\right)\nonumber
\end{eqnarray}
\[8\pi G\rho=-\frac{1}{B}\left(\frac{1}{r^2}-\frac{B'}{rB}\right)
+\frac{1}{r^2},\] in which the prime denotes derivative with
respect to the radial coordinate $r$. On the other hand, the
components of the energy-momentum tensor must satisfy the
gravitational equation [11] $T_{i;k}^{k}=0$. From this equation we
can now obtain a relation between the metric component $A(r)$ and
the energy density $\rho$, $A(r)'=-2\omega A(r)\rho
'/[\rho(1+\omega)]$, which can be immediately integrated to give
$\alpha\rho=A(r)^{-(1+\omega)/(2\omega)}$, with $\alpha$ again an
arbitrary integration constant. Note that for $\omega=-1$, $\rho$
becomes a simple constant, so that the Einstein equations are
straightforwardly solved to produce the well-known static de
Sitter metric. In the case $\omega=-1/3$ we can obtain the
solution [11,16]
\begin{equation}
A(r)=K_1 -\frac{8\pi GK_2}{3\alpha}\sqrt{1+ \frac{3\alpha
K_3}{8\pi Gr^2}}
\end{equation}
\begin{equation}
B(r)=-\left[\left(K_3+\frac{8\pi
Gr^2}{3\alpha}\right)A(r)\right]^{-1} ,
\end{equation}
where $K_1$, $K_2$ and $K_3$ are integration constant which are
arbitrary unless by the condition $K_1 K_3=-1$. This metric shows
an event horizon at \[r=r_h=\left(\frac{3\alpha K_3 K_1^2}{8\pi
GK_2^2} -\frac{8\pi G}{3\alpha K_3}\right)^{-1} ,\] which marks
the transition from a Kleinian-signature metric for $r<r_h$ to an
Euclidean (Riemannian) metric for $r>r_h$. For the entire range of
$\omega$-values in the accelerating interval $-1/3>\omega >-1$, we
have the solution [11]
\[A(r)=K(\omega,\alpha)r^{4\omega/(1+\omega)} \]
\[B=\frac{\omega^2+6\omega +1}{(1*\omega)^2}  ,\]
where
\[K(\omega,\alpha)=\left(\frac{2\pi G(1+\omega)^2
B}{\omega\alpha}\right)^{2\omega/(1+\omega)} ,\;\; -1<\omega<-1/3
\]
Note that in the considered interval $B$ reduces to a simple
dimensionless constant which is negative definite. Besides being
singular at the origin of radial coordinate $r$, the two most
interesting properties of this solution are: (i) it does not show
any event horizon (so that, all world lines should necessarily be
always connected to each other in both the static spacetime and,
in spite of corresponding to an accelerating FRW spacetime, the
cosmological spacetime described by metric (2.4)), and (ii) it has
a Kleinian definite signature (- - + +). The latter property is
consistent with property (i) and with a Lorentzian wormhole
interpretation of the FRW metric with scale factor (2.4) for the
following reason. Let us first introduce the coordinate change
$r=\ell\tan^2(\psi/2)$, where
$\ell=K(\omega,\alpha)^{-(1+\omega)/(4\omega)}$, with the new
half-periodic coordinate $\psi$ running from $\psi=0 (r=0)$ to
$\psi=\pi (r=\infty)$, and then consider a world line for matter
defined by $\theta,\phi$=const., $t=-\beta\psi$, in which $\beta$
is a real constant. The static metric reduces then to the line
element
\[ds^2= -\left[\beta^2\tan^{8\omega/(1+\omega)}(\psi/2)
-B\ell^2\frac{\tan^2(\psi/2)}{\cos^4(\psi/2)}\right]d\psi^2 .\]
Since $B$ is a negative definite constant along the entire
interval $-1<\omega <-1/3$, and $\beta$ is real, this line element
is always timelike along that interval. Thus, an observer moving
on the world line will always have an increasingly negative time
coordinate, i.e. though that observer cannot even reach the point
$\psi=\pi$ and, therefore, can never follow a closed timelike
curve, it will always travel backward in time.

In this paper, we shall concentrate on the particular value
$\omega=-2/3$ which corresponds to a FRW metric that predicts an
accelerating universe which conforms well to supernova results. In
this particular case, the above solution reduces to
\begin{equation}
A(r)=\left(\frac{3\alpha}{23\pi G r^2}\right)^{4} ,\;\; B=-23 ,
\end{equation}
which retains all the properties which we have discussed for the
generic case.

Most recent results on the large scale curvature of the universe
provided by BOOMERanG [19] and other CMB anisotropy experiments
[20] indicate that the universe is flat with a 95 percent
confidence interval. This leaves a 5 percent uncertainty which,
from the qualitative standpoint, tells us that the geometry of the
universe can still be open or closed. On the other hand, along the
development of physics we are used to finally discover that the
apparently most probable simplest situations (circular orbits for
planets and atomic electrons, spherical symmetry, etc) were often
not the real case, but just a good approximation. If flatness is
taken to represent the simplest possible geometry of the universe,
we shall adhere to follow this empirical tendency by choosing for
our accelerating universe with quintessential equation of state
$\omega=-2/3$ a closed geometry, hoping that, like for previous
historical examples, this finally uncovers a richer structure.
Moreover, one should expect that the conclusions on the stability
of the universe against cosmological perturbations we are going to
obtain in the next section for closed geometry would be also
shared by flat geometry. In any event, a detailed consideration of
the cases for flat and open geometries are left for future
research. It is only for the sake of completeness that we shall
finally include the metrical solution for any constant $\omega$ in
the accelerating interval for the flat geometry. In FRW
coordinates it reads
\begin{equation}
ds^2= -dt^2+ \left(\frac{6\pi G(1+\omega)^2
t^2}{\alpha}\right)^{1/[3(1+\omega)]}\left(dr^2+r^2
d\Omega_2^2\right) .
\end{equation}
This line element is valid for all $-1/3>\omega>-1$ and is
associated with a corresponding static metric for any accelerating
value of $\omega$ [11] of which the metric coefficients in Eq.
(2.13) become just the case for $\omega=-2/3$. Therefore, metric
(2.14) and its static counterpart should possess exactly the same
properties as what were discussed before for metrics (2.4) and
(2.13).

\section{Cosmological perturbations}

In this section we shall first briefly review those mathematical
aspects of the perturbative technique developed first by Lifshitz
[14] which will be later used to investigate the stability of our
closed accelerating universe. For a closed FRW spacetime,
perturbations of the four-metric are introduced by
$g_{ab}=g_{ab}+h_{ab}$, with the perturbations satisfying the
gauge $h_{00}=h_{0\alpha}=0$, $a,b,...=0,1,...,3$ and
$\alpha,\beta,...=1,...,3$. It is work remarking that, although we
are working in an isotropic framework with a reference system
which is always synchronous, after this choice of gauge, it is no
longer a comoving system because the perturbations on the spatial
components of the four-velocity can be generally nonzero
[14,22,23]. Perturbation of the Ricci tensor and, hence the
Einstein equation, energy density and velocity can then be derived
(for details see Refs. [21,22]). By taking as the most general
metric perturbation
\begin{equation}
h_{\alpha\beta}=\lambda(\eta)P_{\alpha\beta}
+\mu(\eta)Q_{\alpha\beta}+\sigma(\eta)S_{\alpha\beta}
+\nu(\eta)H_{\alpha\beta} ,
\end{equation}
where the coefficients $\lambda, \mu,\sigma$ and $\nu$ depend only
on the $\eta$ time parameter, and $P_{\alpha\beta}$,
$Q_{\alpha\beta}$, $S_{\alpha\beta}$ and $H_{\alpha\beta}$ are
tensor harmonics which are defined by [22]
\begin{equation}
Q_{\alpha\beta}=\frac{1}{3}\gamma_{\alpha\beta}Q, \;\;
\nabla^a\nabla_a Q^{(n)}=\left(1-n^2\right)Q^{(n)}
\end{equation}
\begin{equation}
P_{\alpha\beta}=\frac{\nabla_{\alpha}\nabla_{\beta}Q}{\ell(\ell+2)}
+Q_{\alpha\beta} ,\;\;
S_{\alpha\beta}=\nabla_{\alpha}S_{\beta}+\nabla_{\beta}S_{\alpha},
\end{equation}
\begin{equation}
\;\; \nabla_a\nabla^a S_b^{(n)}=\left(2-n^2\right)S_b^{(n)},\;\;
\nabla^a S_a^{(n)}=0
\end{equation}
\begin{equation}
\nabla_a\nabla^a H_{cd}^{(n)}=\left(3-n^2\right)H_{cd}^{(n)},\;\;
\nabla^a H_{ab}^{(n)}=0,\;\; H_a^{(n)a}=0 ,
\end{equation}
with $Q^{(n)}$, $S_a^{(n)}$ and $H_{cd}^{(n)}$, respectively, the
scalar, vector and tensor harmonics [21,22], an $n$ an integer
order.

Inserting Eq. (3.1) into the perturbed expressions for the
Einstein equations and the energy density and velocity components,
and using the above definitions, we finally obtain (a prime
denotes differentiation with respect to $\eta$)
\begin{equation}
\lambda''+\frac{2a' \lambda'}{a}
-\frac{1}{3}\ell(\ell+2)\left(\lambda+\mu\right)=0
\end{equation}
\begin{eqnarray}
&&\mu''+\left(3C_s^2+2\right)\frac{a' \mu'}{a}\nonumber\\
&&+\frac{1}{3}\left(3C_s^2+
1\right)\left[\ell(\ell+2)-3\right]\left(\lambda+\mu\right)=0 ,
\end{eqnarray}
\begin{equation}
\sigma'' +\frac{2a'\sigma'}{a}=0
\end{equation}
\begin{equation}
\nu''+\frac{2a'\nu'}{a}+\ell(\ell+2)\nu=0
\end{equation}
and
\begin{equation}
\frac{\delta\rho}{\rho}=
\frac{a(\eta=0)^2}{9a^2}\left\{\left[\ell(\ell+2)
-3\right]\left(\lambda+\mu\right) +3\frac{a'\mu'}{a}\right\}Q
\end{equation}
\begin{equation}
\delta v^{\alpha}
=\frac{P^{\alpha}}{12\left[1-\left(a'/a\right)^2\right]}
\left\{\ell(\ell+2)\mu'+
\left[\ell(\ell+2)-3\right]\lambda'\right\} ,
\end{equation}
for scalar harmonics, and
\begin{equation}
\frac{\delta\rho}{\rho}=0, \;\; \delta
v^{\alpha}=\left[\ell(\ell+2)-3\right]\sigma' S^{\alpha}
\end{equation}
for vector harmonics.

\subsection{scalar harmonics}

The differential equations for the $\eta$-dependent coefficients
$\lambda$ and $\mu$, representing metrical scalar perturbations of
a closed FRW geometry for the scale factor (2.5) can be written as
\begin{equation}
\lambda''+2\tan(\eta/2)\lambda'
-\frac{1}{3}\ell(\ell+2)\left(\lambda+\mu\right)=0
\end{equation}
\begin{equation}
\mu''-
\frac{1}{3}\left[\ell(\ell+2)-3\right]\left(\lambda+\mu\right)=0 ,
\end{equation}
while the perturbations for energy density and velocity components
become
\begin{equation}
\frac{\delta\rho}{\rho}=
\frac{\cos^4(\eta/2)}{9}\left\{\left[\ell(\ell+2)
-3\right]\left(\lambda+\mu\right) +3\tan\eta\mu'\right\}Q
\end{equation}
\begin{equation}
\delta v^{\alpha}
=\frac{P^{\alpha}}{12\left[1-\tan^2(\eta/2)\right]}
\left\{\ell(\ell+2)\mu'+
\left[\ell(\ell+2)-3\right]\lambda'\right\} ,
\end{equation}
where we have consistently taken for the speed of sound
\begin{equation}
C_s= \delta p/\delta\rho=\omega=-2/3,
\end{equation}
which corresponds to a quintessence scalar field with constant
equation of state $\omega=-2/3$ [12].

The differential equations (3.13) and (3.14) still contain some
residual gauge freedom for a complete specification of the choice
of coordinates [14,21,22]. Such an unphysical gauge corresponds in
the present case to the particular solutions
\begin{equation}
\lambda=-\mu= {\rm Const}
\end{equation}
\begin{equation}
\lambda=\ell(\ell+2)\left(\eta+\sin\eta\right) ,\;\;
\mu=-\ell(\ell+2)\left(\eta+\sin\eta\right)+3\sin\eta .
\end{equation}
These solutions will be conveniently subtracted from the general
solutions to Eqs. (3.13) and (3.14). The latter solutions will be
obtained with the help of the auxiliary functions $\xi$ and
$\zeta$ introduced by the substitutions
\begin{equation}
\lambda+\mu=3\sin\eta\int\xi d\eta
\end{equation}
\begin{equation}
\lambda'-\mu'=2\ell(\ell+2)\left(1+\cos\eta\right)\int\xi d\eta
-3\cos\eta\int\xi d\eta +\zeta\cos\eta .
\end{equation}
When substitutions (3.20) and (3.21) are introduced in Eqs. (3.13)
and (3.14), we obtain the new coupled differential equations
\begin{equation}
\xi'+\cot(\eta/2)\xi+\frac{1}{3}\left[1
-\frac{1}{2}\sec^2(\eta/2)\right]\zeta=0
\end{equation}
\begin{eqnarray}
&&\zeta'+\left[\tan(\eta/2)-\tan\eta\right]\zeta\nonumber\\
&&+\left[2\ell(\ell+2)\left(1+\sec\eta\right)-
3\left(1-\tan\eta\tan(\eta/2)\right)\right]\nonumber\\ &&=0 .
\end{eqnarray}
Straightforward manipulations on these equations lead finally to
\begin{equation}
\zeta=-\sec\eta\cot^2(\eta/2)y'
\end{equation}
\begin{equation}
-y''+\cot(\eta/2)y' +\left[\frac{2}{3}\ell(\ell+2)-\frac{1}{4}+
\frac{3}{4}\tan^2(\eta/2)\right]y=0 ,
\end{equation}
where $y=\xi\sin^2(\eta/2)$.

Our task now is to solve Eqs. (3.24) and (3.25). Since obtaining
exact solutions to these equations in closed form is very
difficult, we shall derive approximate solutions in the extreme
cases when $\eta\rightarrow 0$ (i.e. at the beginning of the
accelerating phase) and $\eta\rightarrow\pi$ (i.e. toward the
asymptotic future of the eternally accelerating expansion). In the
first stages of accelerating expansion with $\eta<<1$, Eq. (3.25)
can be approximated up to second order in $\eta$ as
\begin{equation}
-\eta y''+\frac{1}{6}\left(12-\eta^2\right)y'
+\left[\frac{2}{3}\ell(\ell+2)-\frac{1}{4}\right]\eta y=0 .
\end{equation}
At the smallest $\eta$ the solution can in turn be approximated in
terms of Bessel function $J_{\nu}$ in the form
\begin{equation}
y\simeq\eta^{3/2}J_{3/2}\left(\sqrt{\frac{1}{4}-
\frac{2}{3}\ell(\ell+2)}\eta\right).
\end{equation}
As $\eta\rightarrow 0$, we then have $y\propto\eta^{3/2}$. It
follows that for the auxiliary functions $\xi\propto\eta$,
$\zeta\propto-4(1-\eta^{2}/3)$, and hence, since all residual
gauge given by Eqs. (3.10) will vanish as $\eta\rightarrow 0$, one
consistently concludes that the metric and dark energy (density
and velocity components) perturbations all vanish as one
approaches the onset of the accelerating region. Had we chosen for
the Bessel function any of the functions $H_{\nu}$ [23], then all
the above perturbations would be pure imaginary and divergent as
$\eta\rightarrow 0$.

Of greater interest to study the stability of our eternally
accelerating universe is to consider the behaviour of
perturbations in the asymptotic region $\eta\rightarrow\pi$ (i.e.
$t\rightarrow\infty$). Thus, we next look at the solutions to Eqs.
(3.24) and (3.25) as $\eta\rightarrow\pi$. Let us first introduce
the change of time coordinate $x=\tan(\eta/2)$ in Eq. (3.25) which
then becomes \[-\left(1+x^2\right)^2\frac{d^2 y}{dx^2}\]
\begin{equation}
+\frac{2}{x}\left(1-x^4\right)\frac{dy}{dx}+
\left[\frac{8}{3}\ell(\ell+2)-1+3x^2\right]y=0,
\end{equation}
which for large $x$ and even moderate $\ell$ admits an approximate
solution again expressible in terms of the Bessel function
$J_{\nu}$, i.e.
\[y\simeq\sqrt{\cot(\eta/2)}J_{\sqrt{13}/2}\left[\sqrt{8\ell(\ell+2)
-1}\cot(\eta/2)\right].\] As $\eta\rightarrow\pi$ this solution
reduces to:
\begin{equation}
y\propto \left[\cot(\eta/2)\right]^{(1+\sqrt{13})/2} ,
\end{equation}
or when expressed in terms of the auxiliary functions $\xi$ and
$\zeta$,
\begin{equation}
\xi\simeq
A(\ell)\sin^{-2}(\eta/2)\left[\cot(\eta/2)\right]^{(1+\sqrt{13})/2}
\end{equation}
\begin{equation}
\zeta\simeq\frac{1}{4}{\rm sec}\eta\left(1+\sqrt{13}\right)\xi ,
\end{equation}
where $A(\ell)$ is a finite constant that depends only on $\ell$
and whose precise value is not of interest in this paper. The use
of these expressions in Eqs. (3.20) and (3.21) leads finally to
the solutions for the perturbations coefficients
\begin{equation}
\lambda=\frac{A(\ell)}{3+\sqrt{13}}\left[B(\ell)-
12\sin^2(\eta/2)\right]\left[\cot(\eta/2)\right]^{(5+\sqrt{13})/2}
\end{equation}
\begin{equation}
\mu=-\frac{A(\ell)}{3+\sqrt{13}}\left[B(\ell)+
12\sin^2(\eta/2)\right]\left[\cot(\eta/2)\right]^{(5+\sqrt{13})/2}
,
\end{equation}
where $B(\ell)$ is a finite constant given by
\[B(\ell)=\frac{16}{5+\sqrt{13}}\left[1
-\sqrt{13}/8 -\ell(\ell+2)\right] ,\] and we have subtracted the
unphysical gauge associated with the particular solution given by
Eqs. (3.19), i.e. $\lambda=2C_1 \ell(\ell+2)(\eta+\sin\eta)+C_2$,
$\mu=-2C_1
\left[\ell(\ell+2)(\eta+\sin\eta)-3\sin\eta\right]-C_2$, with
$C_1$ and $C_2$ two arbitrary integration constants. Using then
Eqs. (3.15) and (3.16) we finally derive the perturbation in
energy density and velocity components for the quintessence field
which are given by
\begin{widetext}
\begin{eqnarray}
&&\frac{\delta\rho}{\rho}=
\frac{2A(\ell)\cos^4(\eta/2)\left[\cot(\eta/2)\right]^{(5+
\sqrt{13})/2}}{3\left(3+\sqrt{13}\right)\csc^2(\eta/2)}\times\nonumber\\
&&\left\{-4\left[\ell(\ell+2)- 3\right] -12\sin\eta\cot(\eta/2)+
\frac{5+\sqrt{13}}{4}\left[B(\ell)\csc^2(\eta/2)+12\right]\right\}Q
\end{eqnarray}
\begin{eqnarray}
&&\delta v^{\alpha}=
\frac{A(\ell)\cos^2(\eta/2)\left[\cot(\eta/2)\right]^{(3+
\sqrt{13})/2}}{(3+\sqrt{13})\cos\eta}
\left\{\ell(\ell+2)\left[\frac{(5+
\sqrt{13})}{4}\left(B(\ell)\csc^2(\eta/2)+12\right)
-12\sin\eta\left[\cot(\eta/2)\right]^{(2+
\sqrt{13})/2}\right]\right.\nonumber\\
&&\left.-\left[\ell(\ell+2)-3\right]\left[\frac{(5+
\sqrt{13})}{4}\left(B(\ell)\csc^2(\eta/2)+12\right)+
12\sin\eta\left[\cot(\eta/2)\right]^{(2+
\sqrt{13})/2}\right]\right\}P^{\alpha} .
\end{eqnarray}
\end{widetext}
All of the Eqs. (3.32) - (3.35) vanish as $\eta\rightarrow\pi$.
Had we taken any of the two Bessel functions $H_{\nu}$ [23]
instead of $J_{\nu}$ for the solution of the differential equation
(3.28), then we would have finally obtained the same expressions
as (3.32) - (3.35), but with the sign for all $\sqrt{13}$ changed.
These would again vanish as $\eta\rightarrow\pi$, so the
accelerating closed universe resulting from the presence of a
quintessence scalar field with constant equation of state
$\omega=-2/3$ appears to be stable to scalar Lifshitz-Khalatnikov
perturbations.

\subsection{vector and tensor harmonics}

For the case under consideration, perturbations associated with
vector harmonics are described by coefficients that satisfy the
differential equation
\begin{equation}
\sigma''+2\tan(\eta/2)\sigma'=0 ,
\end{equation}
with
\begin{equation}
\frac{\delta\rho}{\rho}=0 ,\;\; \delta
v^{\alpha}=\left[\ell(\ell+2)-3\right]\sigma' S^{\alpha} .
\end{equation}
The solution to Eq. (3.36) can be given in closed form and reads
\begin{equation}
\sigma=C_0 +C_1\left(\frac{3}{4}\eta+\sin\eta
+\frac{1}{8}\sin(2\eta)\right) ,
\end{equation}
where $C_0$ and $C_1$ are arbitrary integration constants. Now,
from Eqs. (3.37) it follows that $\sigma'=0$, so the constant
$C_1$ should be zero too, and hence $\sigma=C_0$. Therefore, as
usual [14,21,22], vector perturbations correspond to unphysical
pure gauge and are hence irrelevant also for the eternally
accelerating universe we are considering in this paper.

Let us next deal with the quite more interesting study of the
gravitational waves associated with the perturbations generated by
tensor harmonics. In our case, the coefficients for such
perturbations are described by the differential equation
\begin{equation}
\nu'' +2\tan(\eta/2)\nu' +\ell(\ell+2)\nu=0 .
\end{equation}
Two cases can now be distinguished. If $\ell=0$, then the solution
for $\nu(\eta)$ is formally identical to that for $\sigma(\eta)$
given by Eq. (3.38), but its interpretation is different. It
physically represents the time evolution of gravitational waves.
We have obtained that, even though the gravitational wave
amplitude does not vanish as $\eta\rightarrow\pi$, neither it grow
as $t\rightarrow\infty$, at which limit $\nu=C_{*}\equiv C_0+3\pi
C_1/4$=Const. The neat effect of the whole evolution from $\eta=0$
to $\eta=\pi$on the zero-mode amplitude is an increase from $C_0$
to $C_{*}$. If $\ell\neq 0$, then by introducing the coordinate
change $z=\sin(\eta/2)$, Eq. (3.39) can written as
\begin{equation}
(1-z^2)\nu'' +3z\nu' +m(m+4)\nu =0,
\end{equation}
in which $m=2\ell$. The solution to this differential equation can
most easily be expressed in terms of ultraspherical (Gegenbauer)
polynomials $C_n^{(3)}$ [23] of odd degree $n$ and reads
\begin{equation}
\nu(\eta)=\cos^{5/2}(\eta/2)C_{2\ell-
1}^{(3)}\left[\sin(\eta/2)\right] , \;\; \ell>0 .
\end{equation}
We can then show that this solution becomes proportional to
$t^{-5/2}$ as one approaches the asymptotic limit $\eta=\pi$.
Thus, these gravitational modes (which all start with vanishing
amplitude at $\eta=0$) become asymptotically suppressed as one
goes to $t=\infty$. We have thereby excluded any unstable growing
modes of the gravitational radiation in the accelerating regime
driven by a quintessence field with constant equation of state
$\omega=-2/3$.

\subsection{No-Hair conjecture}

Now, from the solution (3.38) for $\ell=0$ we obtain at large $t$,
\begin{equation}
\nu_{\omega=-2/3}^{(0)}\simeq C_{*}+\sqrt{\frac{27\alpha}{128\pi
Ga}}C_1 .
\end{equation}
Thus, even though the value of $\nu$ provided by Eq. (3.30) would
be expected to be larger than the corresponding value for the
zero-mode, $\nu_{ds}^{(0)}=C_2$=Const., for de Sitter space [21],
one may always choose the value of the constant $C_{*}, C_1$ and
$C_2$ such that $\nu_{\omega=-2/3}^{(0)}$ became smaller than
$\nu_{ds}^{(0)}$ at sufficiently later times. For $\ell\neq 0$,
from solution (3.41) at large $t$, we also obtain
\begin{equation}
\nu_{\omega=-2/3}^{(\ell)}\simeq
\frac{(2\ell+4)!}{5!(2\ell-1)!}\left(2a\right)^{-5/4} .
\end{equation}
The comparison of this with the corresponding de Sitter expression
[21] $\nu_{ds}^{(\ell)}\simeq A_1+A_2 \exp(-3Ht)$ (where $H$ is
the Hubble constant and $A_1$ and $A_2$ are constants which only
depend on $\ell$) clearly implies that $\nu_{\omega=-2/3}^{(\ell)}
<\nu_{ds}^{(\ell)}$ even at moderate values of time $t$.

Moreover, from the above discussion it follows that any physical
effects driven by the gravitational radiation modes (3.42) and
(3.43) should in any event be very small, since their physical
wavelengths respectively increase with $a^{1/6}$ and $a^{5/12}$.
Thus physical quantities involving at least two derivatives of the
metric are then suppressed asymptotically by powers of the inverse
of these wavelengths. These results appears to be implying a
cosmic "no-hair" theorem [24] for our accelerating closed universe
endowed with a quintessence field with constant equation of state
$\omega=-2/3$. According, furthermore, to our discussion above, it
would rather be a question on the relative values of the constants
involved in the expressions for coefficients $\nu$ whether the
no-hair de Sitter attractor or our no-hair attractor is the final
solution for an accelerating universe. Of course, the model
discussed in this paper is a simple one, so that one should extend
this discussion to include other models with different spatial
geometries and constant or "tracking" equations of state.

\section{Conclusions}

Motivated by the huge impact produced by supernova observations
[1], in this paper we have studied some new characteristics of the
cosmological and static spacetimes of accelerating universes whose
expansion is driven by a slowly-varying quintessence scalar field
with constant equation of state $-1/3>\omega >-1$. Particularizing
at the observationally most favored case $\omega=-2/3$, it was
shown that these spacetimes do not possess any future event
horizon, a property which is not obviously shared by de Sitter
space. The reason that justifies this result is that, even though
closed timelike curves are not permitted to exist in these
spacetimes, world lines traveling backward in time are allowed to
occur on it, so connecting any otherwise causally disconnected
regions, and hence preventing the formation of future event
horizons. This result could have considerable interest for
particle physics and quantum gravity because it manifestly avoids
the serious, perhaps fatal difficulties for string theory (and
actually any quantum field theory that depends on the presence of
particles at infinity) posed by the existence of a future event
horizon [8-10]. This would be just another reason in favor of
preferring quintessence over a positive cosmological constant,
since asymptotically de Sitter space inexorably leads to the
formation of an event horizon.

The stability of the closed space with accelerating expansion
induced by a quintessence field with an equation of state
$\omega=-2/3$ to the cosmological Lifshitz-Khalatnikov
perturbations [14] on the three-sphere has been also studied in
detail. Although there exist more elaborated, covariant methods
for dealing with cosmological perturbations [25], we have followed
here the original Lifshitz-Khalatnikov treatment because of its
greater adequacy to distinguish among the involved physical
effects. It was obtained that, at least for a closed geometry in
the case $\omega=-2/3$ the space is stable to both, the scalar and
tensorial perturbations, and that the damping of small physical
effects induced by the resulting gravitational waves allows one to
conjecture that -much like it happens in asymptotic de Sitter
space- eternally accelerating universes induced by quintessential
fields have no-hair, so becoming final attractors along the
evolution of the universe. It would be the value taken on by the
constants that characterize the amplitude of the gravitational
radiation what would finally decide whether the attractor of de
Sitter universe or the attractors of $\omega>-1$ quintessential
accelerating universes are going finally to dominate and drive the
future cosmological evolution. This is a matter which cannot be
decided in the present paper, but that appears to be decisive to
avoid the above-mentioned severe conflict between accelerating
universe and string theory (or possibly any competing quantum
theory, if there were any). Since our present understanding of any
of the theories involved at this conflict is still too
rudimentary, it is still rather premature to say anything definite
about it.

It is expected that the results obtained in this paper for the
closed geometry can be applied to flat geometry too. In
particular, it appears that the no-hair conjecture be a key
ingredient also for the spatially flat case. Finally, the rather
intriguing implication that in any eternally accelerating universe
driven by quintessence there would be world lines (followed by
light signals and possibly some kind of matter) traveling backward
in time need further consideration. After all, if such lines were
allowed to exist, then the meaning of cosmological evolution
itself should actually require a deep revision.

\acknowledgements

\noindent This work was supported by DGICYT under Research Project
No. PB98-0520.

\end{document}